\documentclass[twocolumn,amsmath,amssymb,floatfix,superscriptaddress,prb]{revtex4-2}
\usepackage[lcgreekalpha]{stix2}
\usepackage{bm,graphicx,url,epsf,color}
\usepackage{amssymb}
\usepackage{amsthm}
\usepackage{xcolor}
\usepackage{comment}
\usepackage{relsize}
\usepackage{xspace}
\usepackage[normalem]{ulem}
\usepackage[super]{nth}
\usepackage{braket}

\graphicspath{{graphics/}}

\usepackage[colorlinks=true,linkcolor=blue,citecolor=blue]{hyperref}

\begin{document}

\title{Resilience of the quantum critical line in the Schmid transition}

\author{Nicolas Paris}
\affiliation{Universit\'e Paris Cit\'e, CNRS,  Laboratoire  Mat\'eriaux  et  Ph\'enom\`enes  Quantiques, 75013  Paris,  France}
\affiliation{Sorbonne Universit\'e, CNRS, Laboratoire de Physique Th\'eorique de la Mati\`ere Condens\'ee, LPTMC, F-75005 Paris, France}
\author{Luca Giacomelli}
\affiliation{Universit\'e Paris Cit\'e, CNRS,  Laboratoire  Mat\'eriaux  et  Ph\'enom\`enes  Quantiques, 75013  Paris,  France}
\author{Romain Daviet} 
\affiliation{Institut f\"ur Theoretische Physik, Universit\"at zu K\"oln, D-50937 Cologne, Germany}
\author{Cristiano Ciuti}
\affiliation{Universit\'e Paris Cit\'e, CNRS,  Laboratoire  Mat\'eriaux  et  Ph\'enom\`enes  Quantiques, 75013  Paris,  France}
\author{Nicolas Dupuis}
\affiliation{Sorbonne Universit\'e, CNRS, Laboratoire de Physique Th\'eorique de la Mati\`ere Condens\'ee, LPTMC, F-75005 Paris, France}
\author{Christophe Mora}
\affiliation{Universit\'e Paris Cit\'e, CNRS,  Laboratoire  Mat\'eriaux  et  Ph\'enom\`enes  Quantiques, 75013  Paris,  France}

\begin{abstract}

Schmid predicted that a single Josephson junction coupled to a resistive environment undergoes a quantum phase transition to an insulating phase when the shunt resistance $R$ exceeds the resistance quantum $h/(4 e^ 2)$. Recent measurements and theoretical studies have sparked a debate on whether the location of this transition depends on the ratio between the Josephson energy and the charging energy. We employ a combination of innovative analytical and numerical techniques, never before explicitly applied to this problem, to demonstrate decisively that the transition line between superconducting and insulating behavior is indeed independent of this energy ratio. First, we apply field-theory renormalization group methods and find that the $\beta$ function vanishes along the critical line up to the third order in the Josephson energy. We then identify a simple fermionic model that precisely captures the low-energy physics on the critical line, regardless of the energy ratio. This conformally invariant fermionic model is verified by comparing the expected spectrum with exact diagonalization calculations of the resistively shunted Josephson junction, showing excellent agreement even for moderate system sizes. 
Importantly, this identification provides a rigorous non-perturbative proof that the transition line is maintained at $R=h/(4 e^ 2)$ for all ratios of Josephson to charging energies. The line is further resilient to other ultraviolet cutoffs such as the plasma frequency of the resistive environment. Finally, we implement an adiabatic approach to validate the duality at large Josephson energy.

\end{abstract}

\maketitle

\section{Introduction}
Understanding the coherence of a macroscopic quantum variable coupled to a dissipative environment is crucial for controlling its dynamics. Quantum circuit technologies offer ways to implement strong environment coupling and operate quantum phase transitions guided by the environment. A paradigm for this physics is the Schmid transition~\cite{Schmid83,Bulgadaev84} which predicts the localization or absence of localization of a quantum particle in a periodic potential depending on the strength of friction. The same model arises when considering the phase across a Josephson junction~\cite{Schoen90} shunted by a resistor, or equivalently a transmission line. When the shunt resistance $R$ exceeds the resistance quantum $R_Q = h/(4 e^2)$, the phase delocalizes and the junction becomes insulating. In contrast, $R < R_Q$ preserves the Josephson coherence, and a quantum phase transition between these two regimes occurs at $R = R_Q$. 
Although the transition was originally predicted to depend on the ratio between the Josephson energy and the charging energy~\cite{Bulgadaev84}, it was later on conjectured~\cite{Guinea85,Fisher85a,Zwerger86}, but not proven, using renormalization group (RG) and duality arguments, that it always lies at $R=R_Q$, irrespective of how large the Josephson energy is compared to the charging energy.

A longstanding belief among physicists has been that this phase transition exists in resistively shunted Josephson junctions. This belief was strengthened by Monte Carlo simulations~\cite{Kimura2004,Werner05,Lukyanov07}, that confirmed the transition and its location. However, the experimental verification of this effect has proven to be notoriously challenging~\cite{Yagi97,Penttilae99,Penttilae01,Kuzmin91,Leger23,subero2023}, despite recent successes in corroborating the transition scenario~\cite{Kuzmin23} and related theoretical works~\cite{houzet2020,Giacomelli23,burshtein2023,yamamoto2024,yeyati2024}. In more recent years, established properties of the transition have been seriously questioned and controversially disputed~\cite{Murani20,Hakonen21,Altimiras23}, casting doubt on the location of the transition~\cite{Masuki22,Yokota23,Sepulcre22,Masuki22a,Daviet23} and even on the correctness of the Schmid localization-delocalization framework for the shunted Josephson junctions.

\begin{figure}
    \centering
    \includegraphics[width=0.45\textwidth]{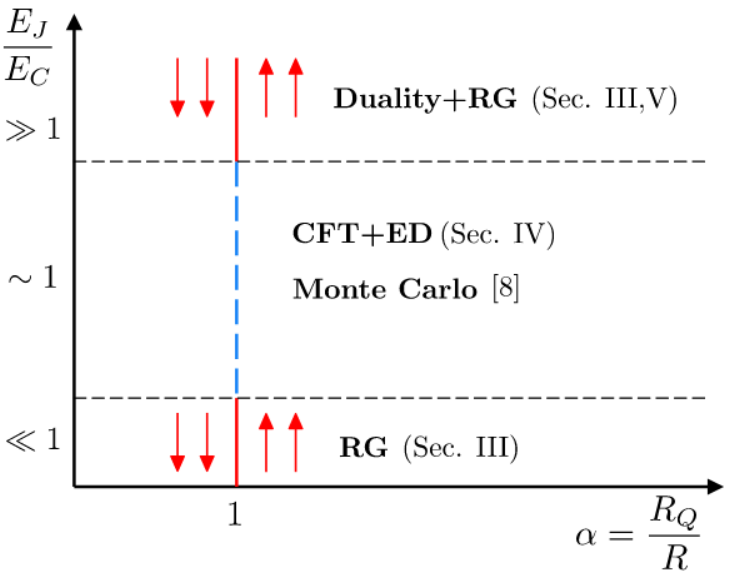}
    \caption{\label{fig:results} Zero temperature phase diagram of the resistively shunted Josephson junction and outline of this paper.    
    The phase transition is located at $\alpha=1$ and follows a straight vertical line as a function of the ratio  $E_J/E_C$. For $E_J\ll E_C$, the RG flow and the onset of the vertical line are obtained from the $\beta$ function~\eqref{eq:beta_final} at third order in $E_J$ and second order in $\alpha-1$, as detailed in Sec.~\ref{sec:fieldRG}. The duality, discussed in Sec.~\ref{Sec:transmon} for $E_J\gg E_C$, allows using the same RG result for the upper part of the phase diagram. Finally, we identify in Sec.~\ref{sec:CFT} the CFT fermionic model describing the entire critical line at fixed $\alpha=1$, which shows excellent agreement with exact diagonalization at already small system size. This is also consistent with the Monte Carlo calculations of Ref.~\cite{Werner05}.}
\end{figure}

This motivates us to revisit thoroughly the Schmid transition for Josephson junctions. Employing rigorous arguments, we demonstrate conclusively that the quantum phase transition indeed occurs when the shunt resistance equals the resistance quantum, $\alpha \equiv R_Q / R = 1$, irrespective of the depth of the Josephson potential. Upon mapping to a boundary sine-Gordon model, we extend RG calculations and obtain that the $\beta$ function vanishes identically at $\alpha = 1$ to third order in the Josephson energy. Subsequently, drawing on a series of conformal field theory (CFT) studies from the nineties~\cite{callan-klebanov1994,callan1994,polchinksi1994}, which identified a conformally invariant solution to the boundary sine-Gordon model at $\alpha = 1$, we show, inspired by Ref.~\cite{Lukyanov07}, that the CFT analysis naturally provides a non-perturbative proof that the transition remains at $\alpha = 1$. This CFT perspective is further reinforced by the excellent agreement between exact diagonalization calculations of the shunted Josephson model and the CFT solution. Finally, employing an adiabatic method, we derive the dual sine-Gordon model in the limit of a deep Josephson potential, corresponding to a transmon regime~\cite{koch2007charge}. The duality~\cite{Schmid83,Bulgadaev84,Guinea85,Fisher85a}, incorrectly criticized in Ref.~\cite{Masuki22}, is crucial for interpreting the independence of the $\alpha = 1$ transition line from the Josephson energy. Our results are summarized in Fig.~\ref{fig:results}.

\section{\label{sec:BSG}Boundary sine-Gordon models}

\begin{figure}
    \centering
    \includegraphics[width=0.35\textwidth]{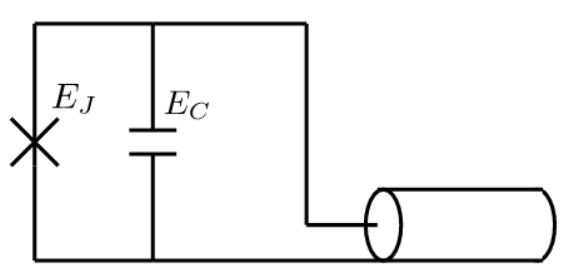}
    \caption{Transmission line terminated by a Josephson junction.}
    \label{fig:initial}
\end{figure}
We consider a Josephson junction, shunted by a resistor modeled as a transmission line, with the Hamiltonian in the charge gauge
\begin{equation}\label{hamiltonian}
    H = 4 E_C (\hat{N} - \hat{N}_0)^2 - E_J \cos \hat \varphi + \sum_q \hbar \omega_q a_q^\dagger a_q .
\end{equation}
$E_C = e^2/2 C$ is the charging energy of the junction capacitance $C$, $E_J$ the Josephson energy (see Fig.~\ref{fig:initial}). The Josephson compact phase $\hat \varphi$ is canonically conjugate to the number of Cooper pairs $\hat{N}$ transferred across the junction. The  transmission line consists of a sum of harmonic modes, with operators $a_q$, which determine the charge displaced to the junction $\hat{N}_0$. The spectral density of the frequencies $\omega_q$ is assumed to be linearly distributed up to sufficiently high frequencies.

Following Ref.~\cite{Morel21}, we connect this system to the boundary sine-Gordon model by employing a unitary transformation akin to a polaron dressing~\cite{ashida2021, Masuki22},
\begin{equation}\label{unitary}
    U = e^{-i \hat \phi_0  \hat{N}}
\end{equation}
where the flux operator $\hat \phi_0$ (in units of $\hbar/2 e$) is conjugate to $\hat N_0$. As detailed in Appendix~\ref{app:gauge-transform}, the unitary transformation removes the linear coupling to $\hat N$ and effectively reduces the mass of $\hat N$  to zero for an infinite transmission line~\cite{ashida2022}. Consequently, $\hat N$ no longer appears in the transformed Hamiltonian, and its conjugate variable $\hat \varphi$ is also eliminated, indicating the loss of charge quantization on the junction due to its coupling to the transmission line. In the limit $E_J \ll E_C$, we obtain a mapping to the boundary sine-Gordon model
\begin{equation}\label{BSG}
    H_{\rm BSG} = \frac{v}{\pi} \int_0^{+\infty} \! \! d x  \left [  \frac{1}{\alpha} (\partial_x \theta)^2 + \frac{\alpha}{4} (\partial_x \phi)^2 \right] 
    -E_J \cos \hat \phi_0,
\end{equation}
where $\phi (x=0) \equiv \hat \phi_0$, supplemented by the UV cutoff $E_C=(2e)^2/2C$ determined by the shunt capacitance $C$ terminating the transmission line. The bulk part of the Hamiltonian involves the flux $\phi (x)$ and its conjugate charge variable $\frac{1}{\pi}\partial_x \theta (x)$ (respectively in units of $\hbar/2 e$ and $2 e$). Eq.~\eqref{BSG} is a reformulation of the transmission line Hamiltonian in the thermodynamic limit, characterized by a series of inductances and capacitances propagating the electromagnetic field with the velocity $v$.

The scaling dimension of the operator $e^{i \hat  \phi_0}$ predicts a quantum phase transition at $\alpha=1$ as the Josephson term is relevant for $\alpha>1$ and irrelevant for $\alpha<1$, corresponding to the superconducting and insulating phases~\cite{Kane1992,Kane1992b,matveev1993}. The marginal line at $\alpha=1$ is proven here rigorously only in the limit $E_J \ll E_C$. The other dual limit $E_J \gg E_C$ is discussed in some detail in Sec.~\ref{Sec:transmon} where an adiabatic approach is employed to derive the low-energy sine-Gordon model. We only quote the final result here and postpone the derivation to Sec.~\ref{Sec:transmon}: the dual Hamiltonian is 
\begin{equation}\label{BSG2}
H_{\rm BSG}^{(2)} = \frac{v}{\pi} \int_0^{+\infty} \! \! d x \left [ \frac{1}{\alpha} (\partial_x \theta)^2 + \frac{\alpha}{4} (\partial_x \phi)^2 \right] - \Delta  \cos 2 \hat \theta_0,
\end{equation}
with $\theta (x=0) \equiv \hat \theta_0$. The UV cutoff is set here by the transmon plasma frequency $\omega_P = \sqrt{8 E_J E_C}$, $\Delta$ is the bandwidth of the lowest band in the spectrum of the isolated transmon. This effective model also predicts a quantum phase transition at $\alpha=1$. In contrast to $H_{\rm BSG}$, the cosine is relevant for $\alpha<1$ and irrelevant for $\alpha>1$. However, the pinning of $\hat \theta_0$ when $\alpha<1$ identifies an insulating phase as for $E_J \ll E_C$ such that the phase diagram is precisely the same in both limits $E_J \ll E_C$ and $E_J \gg E_C$, and the critical $\alpha$ is $1$. The remainder of this paper is devoted to proving that the transition line remains at $\alpha=1$ for all ratios of $E_J/E_C$, as originally conjectured~\cite{Guinea85,Fisher85a,Zwerger86}. 

We stress an important feature~\cite{Morel21} from the sine-Gordon model of Eq.~\eqref{BSG}: it contains an extended phase (flux) $\hat \phi_0$. 
$\hat \phi_0$ is indeed not related to the initial compact phase difference $\hat \varphi$ between the two superconductors, but involves instead continuous photon operators.

\section{Field-theoretic renormalization group}\label{sec:fieldRG}

We explore the RG flow of the boundary sine-Gordon model defined by Eq.~\eqref{BSG}. We focus on the vicinity of the quantum phase transition at $\alpha=1$ and compute the $\beta$-function perturbatively, to third order in $E_J$ and second order in $(\alpha-1)$.  The method is in principle restricted to the regime $E_J \ll E_C$ but, with the dual Hamiltonian of Eq.~\eqref{BSG2}, symmetrically describes the opposite transmon regime of large $E_J$, with the small parameter $\Delta/\omega_P$.

The conventional Wilsonian renormalization group (RG) method proves to be cumbersome at third order. Instead, we implement standard field-theoretical RG techniques, closely following the calculations outlined in Refs.~\cite{Leger23,amit1980renormalisation} for the 2D sine-Gordon model. The scale invariance emerging at criticality underpins the equivalence between the ultraviolet (UV) renormalization and the infrared (IR) scaling at low energies~\cite{le1988critical,delamotte2004hint}.

We consider the Euclidean action associated with the boundary sine-Gordon Hamiltonian~\eqref{BSG}, integrate all bath degrees of freedom and rescale the field $\hat \phi_0 \to \beta_0 \hat \phi_0$,
with $\beta_0^2 \equiv 2\pi/\alpha$.
For simplicity, we drop the hat and the subscript $0$ in $\hat \phi_0$, thus obtaining~\footnote{Alternatively, the action can be written~\cite{Schoen90,Morel21,Daviet23} with a charging energy mass $\sim \omega^2 |\phi (i \omega)|^2/E_C$, reflecting the RC time of the capacitively shunted transmission line. This term is however irrelevant, with dimension $2$, and effectively sets the initial cutoff at energy $E_C$ in the boundary sine-Gordon model.}
\begin{equation}\label{eq:action_BSG}
\mathcal{S}_{\rm BSG}[\phi]=\frac{1}{2}\int\frac{d\omega}{2\pi}|\omega||\phi(i\omega)|^2-\frac{\tilde{E}_J^0}{a\beta_0^2}\int_0^{+\infty} d\tau\, \cos(\beta_0 \phi(\tau))  . 
\end{equation}
The dimensionless bare coupling $\tilde{E}_J^0\sim E_J/E_C$ and the short-time cutoff $a\sim 1/E_C$  define the starting point of the RG flow. 
Their precise values are, however, not needed to compute the RG $\beta$-function.

The renormalisation scheme is the following. To establish the flow equation of $\tilde{E}_J$, we compute the one-particle irreducible (1-PI) two-point vertex $\Gamma^{(2)}(\omega)$ order by order in $\tilde{E}_J^0$. We perform another expansion in $(\alpha-1)$, around $\alpha=1$ where the theory is renormalizable. 
We focus on UV divergences in powers of $\ln a$ encountered in this expansion, as they are sufficient to determine the $\beta$-function.
We find that they can be completely absorbed into a single renormalisation constant $Z_J$ such that the vertex and the renormalised coupling $\tilde{E}_J$ read
\begin{equation}\label{eq:Z}
\begin{split}
    &\tilde{E}_J^0=Z_J\tilde{E}_J , \\
    &\Gamma^{(2)}(\omega,\tilde{E}_J,\kappa)=\Gamma^{(2)}(\omega,\tilde{E}_J^0,\kappa,a) , 
\end{split}
\end{equation}
where we introduce an IR energy scale $\kappa$ (see later Eq.~\eqref{eq:free_prop}) which controls the evolution of the RG flow.
The beta function $\beta_J$, and thus the flow equation of the coupling $\tilde{E}_J$, can be directly 
derived from the renormalisation constant $Z_J$, 
\begin{equation}\label{eq:beta_J}
    \beta_J=-\left.\frac{\partial \tilde{E}_J}{\partial \ln \kappa}\right|_{\tilde{E}_J^0}=\tilde{E}_J\frac{\partial \ln Z_J}{\partial \ln \kappa}.
\end{equation}
We notice that Eq.~\eqref{eq:Z} does not require a field renormalization. Therefore, it proves that $\beta_0$ and the dissipative coefficient $\alpha$ are not renormalized along the RG flow, as expected, since the local impurity cannot renormalize bulk properties~\cite{Kane1992b,Fisher85a}.

In the perturbation calculation based on the action~\eqref{eq:action_BSG}, one faces both IR and UV divergences.  These can be regularized with both short-time $a$ and long-time $\kappa^{-1}$ cutoffs included in the definition of the free propagator,
\begin{equation}\label{eq:free_prop}
\begin{split}
    G_0(\tau,a)&=\int_{-\infty}^{+\infty} \left.\frac{d\omega}{2\pi}\frac{e^{i\omega y}}{\sqrt{\omega^2+\kappa^{2}}} \right|_{y^2=\tau^2+a^2}\\
    &=\frac{1}{\pi}K_0\left(\kappa\sqrt{\tau^2+a^2}\right),
    \end{split}
\end{equation}
where $K_0$ is the modified Bessel function of the second kind.
The behavior of $G_0$ for $\kappa\tau\ll 1$ is used repeatedly in the following,
\begin{equation}
    G_0(\tau,a)\sim -\frac{1}{2\pi}\ln(\kappa^{2}c(\tau^2+a^2)) ,
\end{equation}
where $c=\frac{1}{4}e^{2\gamma}$ with $\gamma$ denoting Euler's constant.

\begin{figure}
    \centering
    \includegraphics[width=0.45\textwidth]{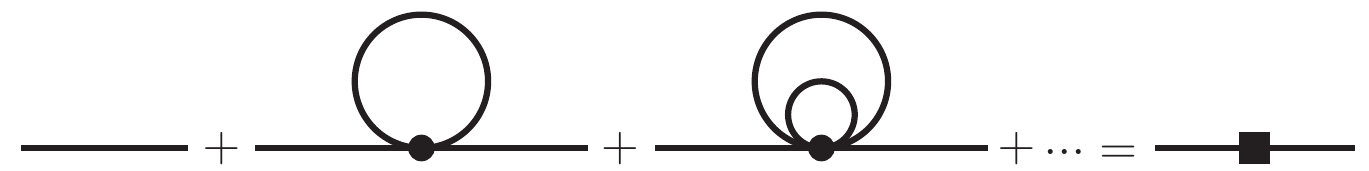}
    \caption{The contributions of all the tadpole diagrams on the left can be resummed in an effective vertex $J$ represented by a square. The numerical prefactors are not shown.} 
    \label{fig:tadpoles}
\end{figure}

The diagrammatic RG theory of the boundary sine-Gordon model presents two main differences with respect to the conventional $\phi^4$-theory. First, the diagrams are most easily expanded in the time domain as the interaction term $\cos\beta_0 \phi$ is local in time and not in frequency. This is consistent with the regularization choice made in Eq.~\eqref{eq:free_prop} as opposed to a frequency cutoff. 
Second, the interaction term $\cos\beta_0 \phi$ intrisically includes an infinite series of powers of $\phi$. Fortunately, systematic resummations can be performed~\cite{amit1980renormalisation}, order by order in $E_J$.
We illustrate this second aspect with first order diagrams, obtained by inserting a single $\cos \beta_0\phi(\tau)$ between $\phi(\tau_1)$ and $\phi(\tau_2)$, as represented in Fig.~\ref{fig:tadpoles}. 
The resulting  infinite series of tadpole diagrams can be resummed  and accounted for by a normal ordering of the cosine together with a renormalization of the coupling~\cite{coleman1975quantum},  
\begin{equation}\label{eq:tadpole}
    \frac{\tilde{E}_J^0}{a}\cos\beta_0\phi\rightarrow \tilde{E}_J^0\kappa\left(\kappa a\right)^{\frac{1}{\alpha}-1}c^{\frac{1}{2\alpha}}:\cos\beta_0\phi:
\end{equation}

\begin{figure}
    \centering
    \includegraphics[width=0.4\textwidth]{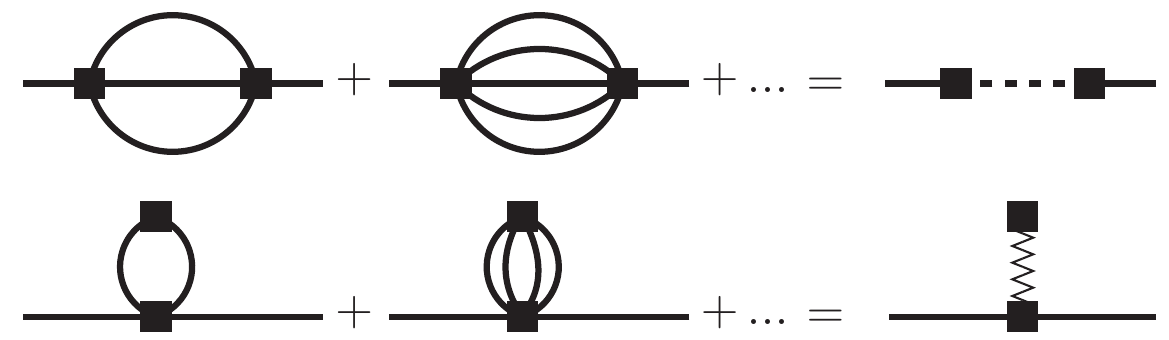}
    \caption{Second-order diagrams contributing to the renormalization of $\Gamma^{(2)}$. On the right-hand side, effective propagators are defined by resumming two subsets of the contributing graphs: those with an odd or even number of inner propagators. The resummation of tadpoles is taken into account with the square vertices.}
    \label{fig:prop}
\end{figure}

In the following, we will use the notation $J=\tilde{E}_J^0\kappa\left(\kappa a\right)^{\frac{1}{\alpha}-1}c^{\frac{1}{2\alpha}}$, which is represented as a square in Figs.~\eqref{fig:tadpoles},~\eqref{fig:prop}. Summing over all tadpole series reorganizes the perturbation theory into an expansion in powers of $J$ (or in the number of squares in diagrams).

It is straightforward yet instructive to compute the $\beta$-function to leading (first) order in $J$, denoted $\beta^{(1)}_J$. The divergence as the short-time cutoff $a$ approaches zero is absorbed into the renormalization constant 
$Z_{J}^{(1)}=( \kappa a)^{1-\frac{1}{\alpha}}c^{-\frac{1}{2\alpha}}$, with $\tilde{E}_J^0=Z_{J}^{(1)} \tilde{E}_J$. The associated $\beta$-function reproduces the scaling dimension of the Josephson coupling: 
\begin{equation}\label{eq:beta_tadpole}
    \beta_{J}^{(1)} =\left(1-\frac{1}{\alpha}\right)\tilde{E}_J.
\end{equation}
A similar argument can be employed to sum over all the second-order diagrams in $J$ shown in Fig.~\ref{fig:prop}. 
The corresponding contribution to the 1PI two-point vertex is
\begin{equation}\label{eq:second_order}
\begin{split}
\Gamma^{(2)}_{2}=&- J^2\beta_0^{-2}\int_{-\infty}^{+\infty} d\tau\, e^{i\omega \tau}\left[\sinh(I(\tau))-I(\tau)\right]\\
&-\left[\cosh(I(\tau))-1\right].
\end{split}
\end{equation}
where $I(\tau)=\beta_0^2G(\tau,a)$. 
The only UV divergence comes from the small $\tau$-behavior of the integrand in~\eqref{eq:second_order}, since 
\begin{equation}\label{eq:small_tau}
\sinh(I(\tau)),\cosh(I(\tau))\sim\frac{1}{2}\left(\frac{1}{\kappa^2 c(\tau^2+a^2)}\right)^{\frac{1}{\alpha}}.
\end{equation}
We expand $e^{i\omega \tau}$ in powers of $\omega$ and analyze each term separately. To zeroth order in $\omega$, the singularities coming from $\cosh I(\tau)$ and $\sinh I(\tau)$ compensate each other exactly. The term proportional to $\omega$ vanishes too as it involves an integral of an odd function of $\tau$. One can show using~\eqref{eq:small_tau} that all the integrals resulting from a higher power of $\omega$ are convergent as long as $\alpha$ is close enough to 1. As a result, $\Gamma_2^{(2)}$ has no UV divergence and does not contribute to $\beta_J$. 
This is expected from simple arguments: the singular $|\omega|$ term cannot be renormalized~\cite{Kane1992b, Fisher85a}, the $\omega^2$ term is irrelevant, and the $\omega$-independent part cannot contribute to the renormalization of the cosine at second order in $\tilde{E}_J^0$ because of the translation symmetry $\phi \rightarrow \phi + \frac{\pi}{\beta_0}$, $\tilde{E}_J^0 \rightarrow -\tilde{E}_J^0$ inherent in the boundary sine-Gordon action and preserved by RG.

Finally, to third order, the $\beta$-function 
is given by
\begin{equation}\label{eq:beta_final}
    \beta_J=\left(1-\frac{1}{\alpha}\right)\left(\tilde{E}_J-\frac{1}{16}\tilde{E}_J^3\right)+\mathcal{O}\left(\tilde{E}_J^5,(\alpha-1)^2\tilde{E}_J^3\right) ,
\end{equation}
as shown in the the Supplemental Material~\footnote{See Supplemental Material at [URL will be inserted by publisher] for the complete derivation of the $\beta$-function at order 3.}.
Remarkably, $\beta_J$ vanishes exactly at $\alpha=1$, providing strong support for the scenario of a vertical line of fixed points in the phase diagram ($\alpha,E_J/E_C$). It contrasts with the case of the spin-boson model, where the third-order term is not vanishing~\cite{Florens2010}. This scenario is further reinforced by the same behavior deduced from the dual Hamiltonian at large $E_J/E_C$. It must be emphasized that, unlike the second-order terms, the cancellation that occurs at third order in $\tilde{E}_J$ cannot be anticipated from straightforward symmetry considerations.
At $\alpha \ne 1$, Eq.~\eqref{eq:beta_final} should be understood as the first two terms of an expansion in $\tilde{E}_J$. Therefore, the apparent zero at $\tilde{E}_J=4$ - not invariant under a change of cutoff prescription - is spurious as it falls outside the validity range of the perturbative approach.  In contrast, we expect $\beta_J$ to be always non-vanishing unless $\alpha=1$ as discussed in the next Section.

\section{Conformal invariance on the critical line}\label{sec:CFT}

\subsection{Boundary CFT}

The vertical transition line at $\alpha=1$, hinted by the results of Sec.~\ref{sec:fieldRG}, can be shown for arbitrary $E_J/E_C$ using CFT arguments pioneered in Refs.~\cite{callan-klebanov1994,callan1994,polchinksi1994}.
The bosonic Hamiltonian of the transmission line of length $L$,
\begin{equation}\label{finite-transmission-line}
H = \frac{v}{\pi} \int_0^L  d x \left [  \frac{1}{\alpha} (\partial_x \theta)^2 + \frac{\alpha}{4} (\partial_x \phi)^2 \right] ,
\end{equation}
exhibits conformal invariance with central charge $c=1$, {\it i.e.} the dispersion relation of modes is linear. The Josephson junction and the capacitance break translational and thus conformal invariance by introducing an energy scale $E^*$. However, for energies well below $E^*$, conformal invariance is restored and the low-energy effective theory becomes a boundary CFT~\cite{Lukyanov07}.

The various boundary CFT associated with the bulk Hamiltonian of Eq.~\eqref{BSG} have been classified for a $2 \pi$-periodic boundary term such as Eq.~\eqref{BSG}. 
For $\alpha <1$ and $\alpha>1$, the only conformally invariant boundary conditions are open (Neumann) or closed (Dirichlet), representing the insulating and superconducting limits (for $c=2$, see {\it e.g.} Ref.~\cite{ashida2024}). 
The case $\alpha=1$ is distinct, revealing a one-parameter family of boundary CFT that smoothly interpolates between Neumann and Dirichlet. It involves two chiral fermion fields, $\Psi = (\psi_1,\psi_2)^T$, related to the original fields by bosonization and living on the unfolded space $[-L,L]$. The boundary CFT is governed  by the Hamiltonian density~\cite{polchinksi1994}
\begin{equation}\label{hamiltonien-density}
h = - i \Psi^\dagger \partial_x  \Psi+ \frac{\pi g}{2}  \, \delta (x) \Psi^\dagger    \tau_x  \Psi,
\end{equation}
where the dimensionless coupling $0<g<1$ interpolates between open ($g=0$) and closed ($g=1$) 
transmission lines, $\tau_x$ is a Pauli matrix.
The mobility $\mu (\omega) \equiv \langle \partial_t\hat \phi_0 (t) \hat \phi_0 \rangle_\omega$ is straightforward to evaluate within this quadratic theory with the result~\cite{Lukyanov07} 
\begin{equation}\label{mobility}
    \mu (\omega) = \mu (0) = \cos^2 \left( \frac{\pi g}{2} \right).
\end{equation}
 The prefactor in the bosonization formula $\psi_1^\dagger (x) \psi_2 (x) \sim e^{i \phi (x)}$ is known analytically~\cite{Guinea85} only in the limit $E_J \ll E_C$. However, the boundary CFT of Eq.~\eqref{hamiltonien-density} remains valid for all values of $E_J/E_C$. Moreover, it more generally applies to any added boundary term to the transmission line that exhibits the periodicity $\hat \phi_0 \to \hat \phi_0 + 2 \pi$.  Except for the two limiting dual cases~\cite{Lukyanov07},
\begin{equation}
    \begin{split}
    \begin{cases}
        g \simeq \frac{\pi}{2 e^{\gamma}} \frac{E_J}{E_C} & \quad (E_J \ll E_C), \\[2mm]
        g \simeq 1 - \frac{2 \Delta}{\omega_P} & \quad  (E_J \gg E_C),
    \end{cases}
    \end{split}
\end{equation}
the link between $g$ and $E_J/E_C$ remains unknown analytically. 
It can, nevertheless, be deduced  from the numerical evaluation of a given physical observable. 

Remarkably, the absence of frequency dependence in the mobility Eq.~\eqref{mobility} is consistent with the lack of physical scales in Eq.~\eqref{hamiltonien-density} and indicates that $g$ is precisely marginal, leading to $\beta_J = 0$ for any value of $E_J/E_C$. 
Such a scale invariance was also observed in NRG calculations for the related model of quantum tunneling in a Luttinger liquid~\cite{Freyn11}.
This conclusively establishes that the boundary between the insulating and superconducting regions forms a vertical line at $\alpha=1$, non-perturbatively extending the findings of Sec.~\ref{sec:fieldRG}.

\subsection{Many-body spectrum at criticality}

The aforementioned arguments at $\alpha=1$ arguably hinge on the emergence of the marginal boundary CFT at low energy. We further identify it by comparing the energy spectrum of the shunted Josephson junction, obtained through exact diagonalization (ED) of Eq.~\eqref{hamiltonian}, with the CFT analytical prediction. The (rescaled) spectrum of the boundary CFT at $\alpha=1$ has been derived in Ref.~\cite{Hasselfield06}. It is given by
\begin{equation}\label{eq:cft-bands}
    \frac{E_\mathrm{cft}(\nu)}{\hbar\delta}=(n+\lambda(\mu,\nu))^2 + p -\frac{1}{24},
\end{equation}
where $n$ is an integer, 
$\mu$ is the mobility given in Eq.~\eqref{mobility},
\begin{equation}
    \lambda(\mu,\nu)=\frac{1}{2\pi}\cos^{-1}\left(\sqrt{\mu}\cos(2\pi \nu)\right),
\end{equation}
and $p = \sum_i i \, p_i$, where $p_i$ is the integer number of excitations in the $i^{\rm th}$ mode. 
Here, $\delta$ represents the typical level spacing, or free spectral range. The spectrum depends on the quasicharge $\nu$ and forms bands as a consequence of the Josephson potential $2 \pi$ periodicity. Remarkably, despite strong interactions, the spectrum coincides with the unperturbed photon spectrum of the transmission line with an additional decoupled mode of energy $\sim (n+\lambda)^2$. These excitations are nevertheless many-body combinations of the original photons and the scattering of incident photons has a strong inelastic part~\cite{houzet2023}.

The ED band spectra of the Hamiltonian~\eqref{hamiltonian}, representing the shunted Josephson junction, have been numerically calculated for various $\alpha$ in Ref.~\cite{Giacomelli23}. A critical behavior was already identified at $\alpha=1$ (regardless of $E_J/E_C$) with a marked low-energy scale invariance. Notably, this scale invariance is consistent with the absence of frequency corrections to the mobility [Eq.~\eqref{mobility}], a consequence of the marginal interaction. In addition, it suggests that the CFT description may apply even to relatively small systems.

\begin{figure*}[t!]
  \centering
  \includegraphics[width=1.0\textwidth]{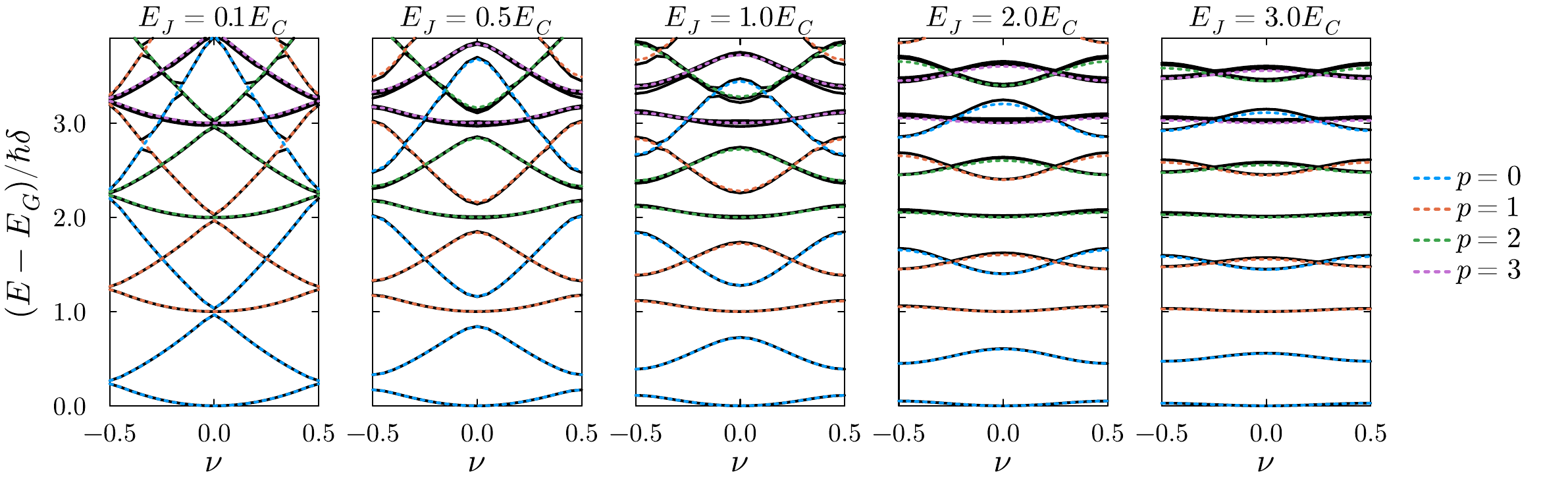}
  \caption{Band spectra at the critical point $\alpha=1$ for different values of $E_J/E_C$. Black lines are the exact diagonalization results obtained with free spectral range $\hbar\delta=0.5E_C$ and plasma frequency $\omega_p=4E_C/\hbar$. Colored dotted lines on top are the CFT result~\eqref{eq:cft-bands}, with the mobility $\mu$ obtained by fitting the first band. Different colors correspond to different \textit{replicas}, labeled by the integer $p$.}
  \label{fig:ED_bands}
\end{figure*}

\subsection{Exact diagonalization}

We numerically solve by ED the spectrum at $\alpha=1$ of the charge gauge Hamiltonian~\eqref{hamiltonian}. Following Ref.~\cite{Giacomelli23}, we  model the finite-size transmission line resonator as a discrete array of unit cells with inductance $\ell$ and capacitance $c$. This introduces a UV cutoff of the mode frequencies given by the transmission line plasma frequency $\Omega=\sqrt{2/\ell c}$ (not to be confused with the transmon plasma frequency $\omega_P$). With this modeling, which is closer to what is experimentally realized in Josephson junctions arrays~\cite{Leger2019,Kuzmin23}, only frequency modes with energies well below $\Omega$ exhibit an equally spaced energy structure and match well with the spectrum of Eq.~\eqref{finite-transmission-line}, with the spectral range $\delta = \pi v/L$ and the velocity $v=\sqrt{1/\ell c}$.

We diagonalize Eq.~\eqref{hamiltonian} for different quasi-charges $\nu$. A convenient basis for a fast convergence of the excited state is chosen by moving to the \textit{polaron frame}, as detailed in Appendix ~\ref{app:exact-diag}. The first band $(E_1(\nu)-E_1(0))/(\hbar\delta)$ is fitted to the CFT prediction Eq.~\eqref{eq:cft-bands}, with $n=0$ and $p=0$, to obtain the mobility $\mu$. Once $\mu$ is estimated, the CFT prediction of Eq.~\eqref{eq:cft-bands} is directly compared to the excited band spectrum in Fig.~\ref{fig:ED_bands}. For different values of $E_J/E_C$, the agreement is excellent between the ED calculation in black lines and the CFT predictions represented as colored dotted lines. The different colors refer to the {\it replicas} of the spectrum labeled by the non-negative integer $p$. There is also a perfect agreement on the degeneracies of bands. The comparison is remarkable at low energies while, as expected, small differences appear at higher energies.

\begin{figure}[t!]
  \centering
  \includegraphics[width=\columnwidth]{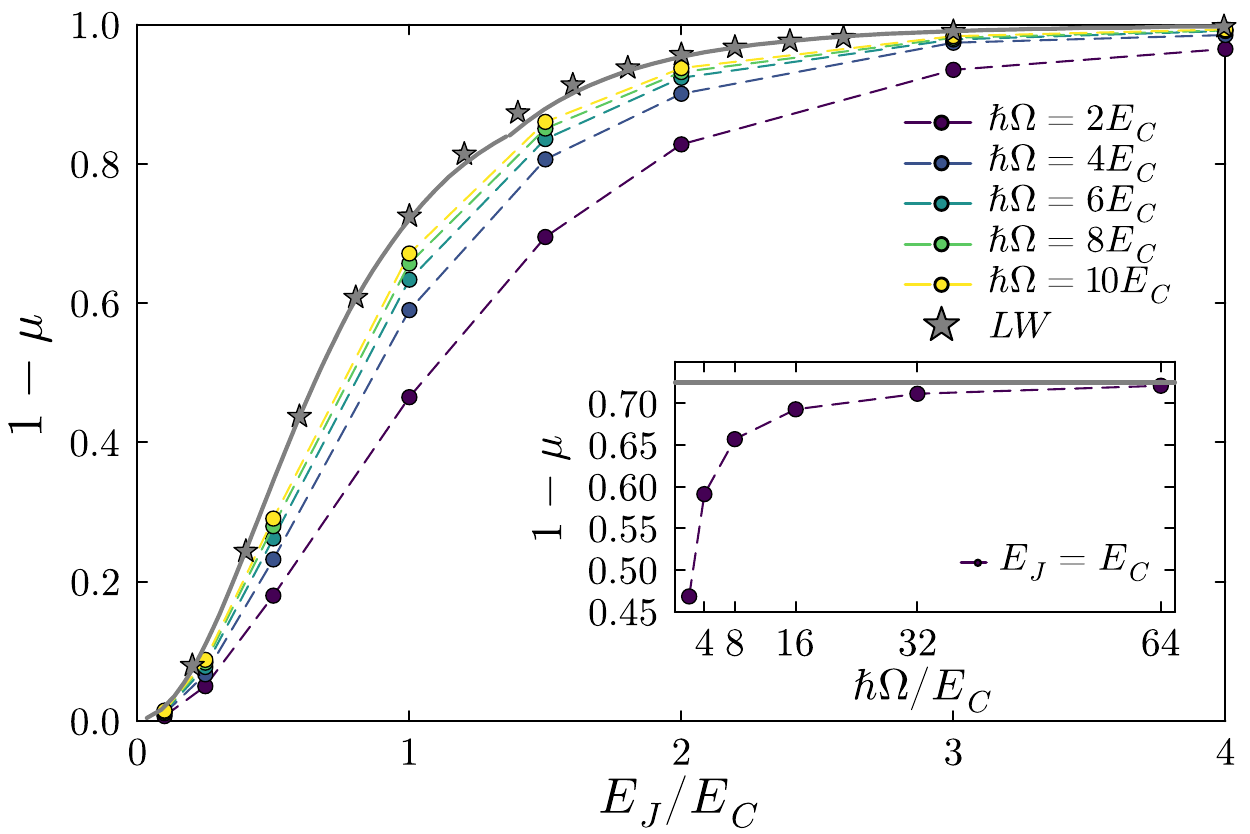}
  \caption{Mobility $\mu$ for different values of $E_J/E_C$ and of $\Omega$, obtained with a fit of the first numerical band with formula~\ref{eq:cft-bands}. The grey stars are the numerical data taken from~\cite{Lukyanov07} and the grey lines are the asymptotic expressions of the same work. In the inset the convergence to the value obtained in~\cite{Lukyanov07} for $E_J=E_C$ is shown. For all the points $\hbar\delta=0.5E_C$.}
  \label{fig:ED_fitted_mob}
\end{figure}

The fitting of the ED lowest band to extract the mobility can be reproduced for various values of $E_J/E_C$ and of $\Omega$. The result, shown in Fig.~\ref{fig:ED_fitted_mob}, nicely interpolates between zero and unit mobility as a function of $E_J/E_C$. We note that $E_J/E_C =4$ is already very close to a vanishing $\mu$ as expected from the exponential scaling in the transmon regime.
The mobility also decreases with the plasma frequency $\Omega$ of the transmission line. In Fig.~\ref{fig:ED_fitted_mob}, we also plot(grey stars and lines) the results obtained by Monte Carlo~\cite{Lukyanov07} in the scaling limit $\Omega/E_C = +\infty$. Our data follow the same trend, approaching this result when increasing $\Omega/E_C$. 
The convergence towards the scaling limit, in very good agreement with the Monte-Carlo result~\cite{Lukyanov07}, is shown for larger values of $\Omega$ in the inset of Fig.~\ref{fig:ED_fitted_mob} for $E_J/E_C=1$.

\begin{figure*}[t!]
  \centering
  \includegraphics[width=1.0\textwidth]{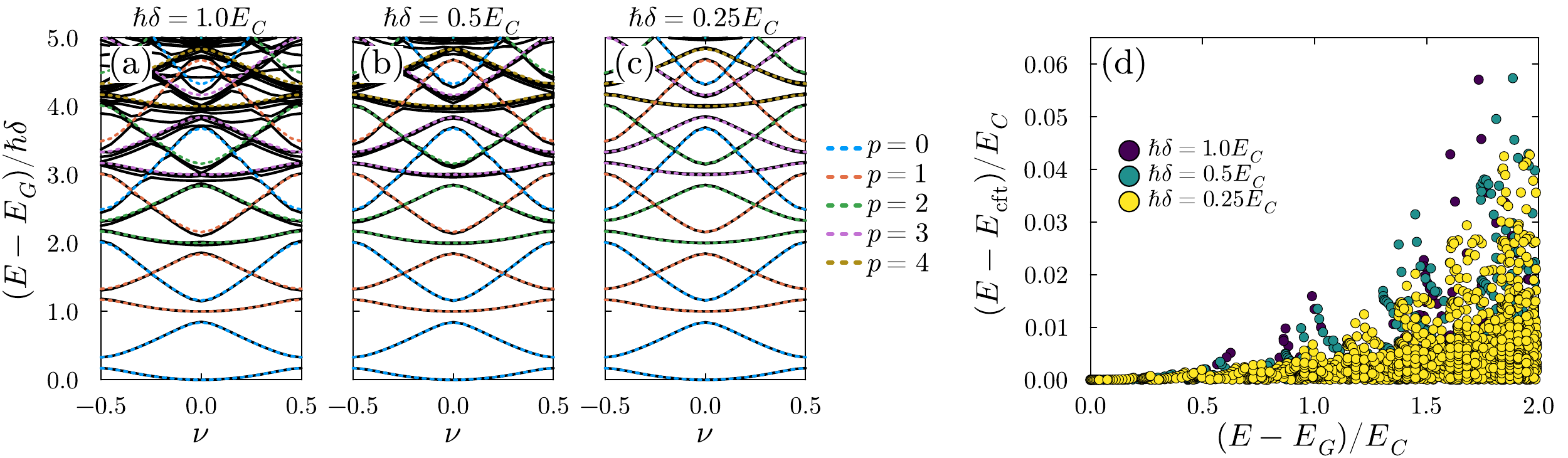}
  \caption{(a)-(c) Size dependence of the band spectrum at the critical point for $E_J=0.5E_C$ and $\hbar\Omega=4E_C$. The black lines are the numerical results, the colored dashed lines the CFT prediction. The number of environmental modes for the different system sizes is respectively 7, 13, 26. (d) Difference between the CFT spectrum and the exact ones for the three system sizes. All the numerically computed energies, for different quasi-charges, are shown. In particular, for the different system sizes, the numbers of energy bands falling in this energy range are respectively 10, 30, 200.}
  \label{fig:ED_size_dep}
\end{figure*}

The agreement with the CFT spectrum is further investigated in Fig.~\ref{fig:ED_size_dep}. In Figs.~\ref{fig:ED_size_dep} (a)-\ref{fig:ED_size_dep}(c) we show the rescaled band spectra for three different system sizes, i.e. different values of $\delta$, while keeping $E_J$ and $\Omega$ fixed. We observe that the number of levels correctly fitting the CFT spectra increases with the system size.
 
In particular, the level degeneracies increasingly align with the CFT predictions. To better highlight the differences, the energy difference between ED and CFT is shown in Fig.~\ref{fig:ED_size_dep}(d), plotted as a function of energy for different system sizes. Although the agreement remains very strong even at high energies, it clearly confirms, with a weak dependence on the system size, that the energy scale $E^*\sim E_C$ marks the threshold below which conformal invariance is restored and the CFT description applies.

Our ED calculations clearly demonstrate the validity of the CFT model at  $\alpha=1$. This is particularly remarkable since our modeling of the shunted junction includes a detailed microscopic description, yet yields excellent agreement with the universal CFT predictions, even for small system sizes. This also bolsters the field theory predictions, in particular the insulator-superconductor phase transition, based on the sine-Gordon model and applied to the resistively shunted Josephson junction.

\section{Transmon limit and dual model} \label{Sec:transmon}

The duality between the two limiting cases $E_J \ll E_C$ and $E_C \ll E_J$, where the conjugate flux and charge variables are simply interchanged $\hat \theta_0 \leftrightarrow \hat \phi_0$, predicts a quantum phase transition in both limits at $\alpha=1$. As noted in the original works~\cite{Guinea85,Fisher85a,Zwerger86}, this supports the scenario of a constant $\alpha=1$ transition line as the most natural one but not the only one. 
Conversely, the breaking of duality claimed in Ref.~\cite{Masuki22}, would strongly challenge the $\alpha=1$ vertical transition line.

The dual picture in the transmon regime of $E_J \gg E_C$ has been discussed extensively in the literature~\cite{Schmid83,Bulgadaev84,Schoen90}: The flux variable (or superconducting phase) $\hat \phi_0$ is most of the time trapped into the $2 \pi$-periodic minima of the deep Josephson potential. The hopping between degenerate minima occurs via instantons with an exponentially small energy
\begin{equation}
    \Delta = 16 E_C \sqrt{\frac{2}{\pi}}\left(\frac{E_J}{2E_C}\right)^{\frac{3}{4}}e^{-\sqrt{8E_J/E_C}}
\end{equation}
and generates a new cosine perturbation (instanton hopping) involving the field dual to $\hat \phi_0$. Adding dissipation gives a boundary sine-Gordon model dual to the one at $E_J \ll E_C$ with $\alpha \to 1/\alpha$.

Starting with the charge gauge Hamiltonian from Eq.~\eqref{hamiltonian}, we now rigorously derive the dual model using an adiabatic approach, similar to the envelope function method employed in semiconductors~\cite{bastard1982} and twisted moiré materials~\cite{bistritzer2011moire,mao2023}. In Eq.~\eqref{hamiltonian}, $\hat N$ and $\hat\varphi$ are fast-evolving variables, with the plasma characteristic frequency $\omega_P$, whereas the dynamics of the slow-moving displaced charge $\hat N_0$ is controlled by $\Delta$.

We first freeze the quantum fluctuations of  $\hat N_0$ and replace it with a classical c-number $N_0$. We thus obtain the Cooper pair box Hamiltonian
\begin{equation}\label{CPB}
    H_{\rm CPB} = 4 E_C (\hat{N} - N_0)^2 - E_J \cos \hat \varphi.
\end{equation}
Its properties have been studied in detail in the context of circuit quantum electrodynamics, notably in the transmon regime $E_J \gg E_C$ where charge noise is minimized~\cite{koch2007charge}. The eigenstates of Eq.~\eqref{CPB} form Bloch bands~\cite{likharev1985theory}, periodic with $N_0$, and noted as Bloch wavevectors $|u^n_{N_0}\rangle$ where $n$ is the band index. For $E_J \gg E_C$, the bands are weakly dispersive and centered around the eigenenergies of the harmonic oscillator $\omega_P (n+1/2)$ as detailed in appendix~\ref{app:transmon}. The lowest band in particular has energy $E_0 (N_0) = - E_J + \hbar \omega_P/2 - \Delta \cos( 2 \pi N_0)$.
\begin{figure}
    \centering
    \includegraphics[width=0.35\textwidth]{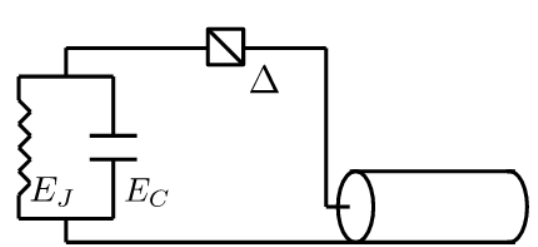}
    \caption{Equivalent circuit in the transmon regime: the transmission line is connected to the charge dissipation device in series with the effective $LC$-oscillator that accounts for the Josephson junction.}
    \label{fig:dual}
\end{figure}
Assuming that the quantum evolution of $\hat N_0$ is slow, $\Delta \ll  \omega_P$, we look for an eigenstate of the full Hamiltonian Eq.~\eqref{hamiltonian} in the form
\begin{equation}\label{eq:ansatz}
    |\psi\rangle =\sum_{N_0,n} f_n(N_0)|u^n_{N_0}\rangle \otimes | N_0 \rangle
\end{equation}
where $| N_0 \rangle$ is an eigenstate of $\hat N_0$ in the transmission line. Injecting this ansatz into the Schr\"odinger equation, we find a set of coupled integral equations verified by the functions $f_n(N_0)$. Details of this calculation are left to appendix~\ref{app:duality}. Interestingly, we find that these equations, in the limit $E_J \gg E_C$, can be mapped onto the simple circuit illustrated in Fig.~\ref{fig:dual}: the transmission line is closed by a non-linear element, with the Hamiltonian $-\Delta \cos ( 2 \pi \hat N_0 ) $, in series with the capacitance $C$ and the Josephson inductance $L_J \equiv (4e^2E_J)^{-1} $. The non-linear term comes solely from the lowest band of the isolated transmon. With $C$ and $L_J$, we nevertheless keep for the moment the trace of the excited bands but neglect their dispersions. We shall later complete the projection to the lowest band. We note that $\Delta$ is exponentially small in the limit $E_J\gg E_C$. However, it cannot be neglected since it is relevant under RG when $\alpha <1$. A similar circuit was previously employed in Ref.~\cite{houzet2023} to describe the regime $\alpha \ll 1$. 

We finally diagonalize the circuit, except for the non-linear element $\sim \Delta$, and find the effective Hamiltonian
\begin{equation}\label{eq:hamiltonian_dual2}
      \tilde{H}_{\rm dual}=\sum_q \hbar \omega_q c_q^\dagger c_q - \Delta \cos 2 \hat \theta_0
\end{equation}
with new bosonic fields $c_q$ for the transmission line shunted by the LC circuit. We introduce the variable $\hat \theta_0 = \pi \hat N_0$ with the mode expansion
\begin{equation}\label{eq:charge_mode}
    \hat \theta_0 =  \sqrt{\frac{\alpha \delta}{2}}  \sum_q \frac{1}{\sqrt{\omega_q}} \left( \frac{1}{1-\frac{i\tau_L\omega_q}{1-\omega_q^2/\omega_P^2}}c_q+h.c.
 \right).
\end{equation}
Two characteristic frequencies emerge: the plasma frequency $\omega_P$ and the $LR$-frequency $\omega_L=1/\tau_L=R/L_J$. Since $\omega_P\ll \omega_L$, the plasma frequency emerges as the natural UV cutoff~\cite{Morel21,houzet2020} below which environment photons do not excite the transmon out of the lowest energy band. We thus eliminate frequencies below $\omega_P$ in Eq.~\eqref{eq:charge_mode} to be consistent with a projection to the lowest band. The fluctuations of the vertex operator $e^{2 i \hat \theta_0}$ are then critical in the long-time limit, $t \gg 1/\omega_P$, with scaling dimension $\alpha$, which justifies the mapping to the dual sine-Gordon Hamiltonian Eq.~\eqref{BSG2} announced in Sec.~\ref{sec:BSG}.

As previously noted~\cite{Lukyanov07}, the dual sine-Gordon Hamiltonian of Eq.~\eqref{BSG2} can be tested at finite temperature $T$. In the regime where $\Delta \ll T \ll \omega_P$, the tunneling events between the minima of the Josephson potential occur incoherently and can be described using rate equations. By employing a WKB approach to evaluate the single instanton tunneling rate, the mobility  in this incoherent regime has been evaluated~\cite{weiss1985}. It coincides precisely with an evaluation using Eq.~\eqref{BSG2} while incorporating the cutoff $\omega_P$.

\section{Conclusions}

We have rigorously demonstrated in this paper that a critical vertical line in the plane $(\alpha=R_Q/R,E_J/E_C)$ separates the phase diagram between an insulating and a superconducting phase for $\alpha$ respectively smaller or larger than $1$. We used a field-theoretical approach to compute the RG-$\beta$ function to third order in $E_J$ in the vicinity of $\alpha=1$. We found in particular that it vanishes identically at $\alpha=1$, indicating that the transition line starts vertically. We also established more thoroughly the duality in this model at large Josephson energy by using an adiabatic approach where the transmon exhibits a slow quantum fluctuating offset charge from the resistive environment. The duality shows that the critical line behaves symmetrically at small and large $E_J/E_C$, which is again consistent with a vertical line.

Building upon Ref.~\cite{Lukyanov07}, we obtained a non-perturbative proof for the location of the transition line. Previous CFT studies on the boundary sine-Gordon model have derived a one-parameter family of fermionic models describing the IR fixed point at $\alpha=1$. This fermionic boundary CFT is identified with the transition line for the resistively shunted Josephson junction. It is exactly marginal, i.e., exhibits no energy dependence, and thus describes a line of fixed points parametrized by the ratio $E_J/E_C$, always at $\alpha=1$. This interpretation is further corroborated by the excellent agreement between the CFT many-body spectrum at different $E_J/E_C$ and exact diagonalization calculations.

We emphasize that the CFT studies from the 1990s~\cite{callan-klebanov1994,callan1994,polchinksi1994} already contained the essence of the non-perturbative proof discussed in this paper. However, the proof was still resting on the two crucial but reasonable assumptions (i) conformal invariance is restored at sufficiently low-energy, and (ii) the identified marginal CFT possesses all the correct symmetries but it is not possible to track the renormalization flow from high to low energies to fully connect it to the original model. Our numerical confirmation bridges this gap by identifying the finite-size spectra of CFT and ED.

One crucial aspect of this phase transition is that it is largely independent of the local details of the Josephson junction but only depends on the resistance of the bulk environment. Seen from the 1D transmission line environment, the only relevant feature of the Josephson junction is its $2 \pi$ periodicity with the boundary flux $\hat \phi_0$. Any other local model exhibiting the same periodicity will undergo a phase transition at $R=R_Q$, regardless of the local microscopic details. The transmission line does not need to be purely ohmic, as long as the density of modes is linear at sufficiently low energy. This might change the UV cutoff of the model but not its critical properties.

Let us finish this discussion with an interesting physical picture for understanding duality at the circuit level, illustrated as Fig.~\ref{fig:initial} for $E_J \ll E_C$ and Fig.~\ref{fig:dual} for $E_J \gg E_C$. In Fig.~\ref{fig:initial}, the transmission line is closed only by a capacitance if we set $E_J=0$. This capacitance behaves like an open circuit at small frequency which allows the flux $\hat \phi_0$ to fluctuate maximally, and potentially disrupt the Josephson trapping, whereas the charge variable $\hat \theta_0$ is quenched. Conversely, in Fig.~\ref{fig:dual} the line is shunted at small frequency by the Josephson inductance (for $\Delta=0$), which behaves as a short circuit, causing $\hat \phi_0$ to be quenched and $\hat \theta_0$ to fluctuate strongly.

Finally, we expect the exact marginality at $\alpha=1$  to have measurable consequences in experiments, resulting in a weak sensitivity to the size of the resistive environment, as well as to frequency, temperature, or generally energy dependencies.

\begin{acknowledgments}
We are grateful to C. Altimiras,  S. Florens, L. Glazman, P. Joyez, H. le Sueur and W. Zwerger   for fruitful discussions.
\end{acknowledgments}

\appendix

\section{Gauge transform}\label{app:gauge-transform}

Starting with the Hamiltonian Eq.~\eqref{hamiltonian} in the charge gauge, we expand $(\hat{N} - \hat{N}_0)^2 $ and change variables for the transmission line,
\begin{equation}\label{free-modes}
 4 E_C    \hat{N}_0^2 + \sum_q \hbar \omega_q a_q^\dagger a_q= \sum_q \hbar \omega_q b_q^\dagger b_q  , 
\end{equation}
while keeping the same density of states. The new bosonic operators $b_q$ diagonalize the transmission line closed on one end by the capacitance $C$~\cite{kaur2021}.
The resulting displaced charge is~\cite{Morel21,houzet2023}
\begin{equation}\label{N0}
\hat N_0 = \frac{1}{\pi} \sqrt{\frac{\alpha \delta}{2}}  \sum_q \left( \frac{-i  \sqrt{\omega_q} \tau}{1- i \omega_q \tau} b_q + h.c.  \right)
\end{equation}
with the RC time $\tau = RC = \hbar \pi/4 \alpha E_C$ and the spectral range $\delta = \pi v/L$,
reflecting the distribution of electromagnetic modes with velocity $v$ and size $2 L$. The conjugate flux operator $\hat \phi_0$ (in units of $2 e/\hbar$) is given by
\begin{equation}\label{phi0}
\hat \phi_0 =  \sqrt{\frac{2 \delta}{\alpha}}  \sum_q \frac{1}{\sqrt{\omega_q}}\left( \frac{1}{1- i \omega_q \tau} b_q + h.c.  \right)
\end{equation}
such that  $[\hat \phi_0,\hat N_0]=i$. The unitary operator Eq.~\eqref{unitary} applied onto the Hamiltonian, $H' = U H U^\dagger$,
completely removes the Cooper pair charge $\hat N$ from the Hamiltonian. This disappearance of $\hat N$, associated with the infinite capacitance of the entire circuit, is strictly valid~\cite{Giacomelli23} only for $L\to +\infty$. Moreover, $U \,  \cos \hat\varphi \, U^\dagger = \cos (\hat \varphi - \hat \phi_0)$. The compact Josephson phase $\hat \varphi$ becomes a constant of motion and can be absorbed into the zero mode of $\hat \phi_0$. After the gauge transform and a lengthy but straightforward computation, we obtain the Hamiltonian
\begin{equation}\label{hamiltonian2}
    H' = U H U^\dagger = \sum_q \hbar \omega_q b_q^\dagger b_q - E_J \cos \hat \phi_0
\end{equation}
At this point, the Hamiltonian~\eqref{hamiltonian2} has been derived from Eq.~\eqref{hamiltonian} without making any approximation. The mapping to the boundary sine-Gordon model is justified in the limit $E_J \ll E_C$. The correlation function for the vertex operator $e^{i \hat \phi_0}$ is computed using the mode expansion of Eq.~\eqref{phi0}. It is critical in the long-time limit, $t \gg \hbar/E_C$, with scaling dimension $1/\alpha$, and thus maps onto the boundary sine-Gordon model given as Eq.~\eqref{BSG} in the main text.

\section{\label{app:transmon}Isolated transmon}
Let us recall the main features of the transmon regime before considering an additional transmission line. The Schrödinger equation associated to the Hamiltonian of a Josephson junction with an external bias $N_0$ , 
\begin{equation}\label{eq:ham_JJ}
H_{N_0}=4E_C(\hat{N}-N_0)^2-E_J\cos \hat{\varphi} ,
\end{equation}
reduces to Mathieu's equation~\cite{wilkinson2018approximate}. Its solutions are given by~\cite{dalibard2013college}
\begin{equation}\label{eq:Wannier}
    u_{N_0}^n(\varphi)=\sum_{p\in \mathbb{Z}}W_n(\varphi-2\pi p)e^{-iN_0(\varphi-2\pi p)} ,
\end{equation}
where $W_n$ denotes the $n^{th}$-Wannier function. The sum simply ensures that the solution is periodic in $\varphi$. When $E_J$ is larger than any other energy scale, the system tends to pin $\varphi$ to $0$ to minimize its energy. In this regime, the Wannier functions can be approximated by the wavefunctions of a harmonic oscillator:
\begin{equation}\label{eq:Hermite}
    W_n(\varphi)\simeq c_n \left(\frac{E_J}{8E_C}\right)^{1/8}H_n\left(\left(\frac{E_J}{8E_C}\right)^{1/4}\varphi\right)e^{-\left(\frac{E_J}{8E_C}\right)^{1/4}\frac{\varphi^2}{2}}
\end{equation}
where $c_n$ is a normalization constant and where $H_n$ is the $n^{th}$-Hermite polynomial. In some sense, the cosine interaction of the Hamiltonian~\eqref{eq:ham_JJ} is simply replaced by its quadratic expansion $1-\varphi^2/2$: the Josephson element becomes an effective inductance.
The associated spectrum is organized in bands that are parameterized by an index $n$ and the charge bias $N_0$~\cite{koch2007charge}:
\begin{equation}\label{eq:energy_level}
\begin{split}
    &E_n(N_0)=E_n-\frac{\epsilon_n}{2}(1+\cos(2\pi N_0)) , \\
    &E_n=-E_J+\omega_P(n+\frac{1}{2})-\frac{E_C}{12}(6n^2+6n+3) , \\
    &\epsilon_n=(-1)^nE_C\frac{2^{4n+5}}{{n!}}\sqrt{\frac{2}{\pi}}\left(\frac{E_J}{2E_C}\right)^{n/2+\frac{3}{4}}e^{-\sqrt{8E_J/E_C}}  , 
\end{split}
\end{equation}
The low-energy spectrum presents some very interesting features: the bands are separated one from another by $\omega_P$, which is very large in the transmon regime, and are characterized by a cosine charge dissipation. For the lowest energy band, the amplitude of the dissipation is given by $\Delta=\epsilon_0/2$.

\section{Dual Hamiltonian\label{app:duality}}

We compute the dual Hamiltonian emerging in the transmon limit where  $E_J\gg E_C$.
The Schrödinger equation with the ansatz~\eqref{eq:ansatz} takes the form $H|\psi\rangle=E|\psi\rangle$ where $H$ is given by equation~\eqref{hamiltonian}. Once projected onto $\langle \varphi|\otimes\langle N_0|$, it yields
\begin{align}
    &\sum_m\left[E_m(N_0)-E\right]f_m(N_0)u_{N_0}^m(\varphi)\nonumber\\
    &+\sum_{m,N_0'}f_m(N_0')u_{N_0'}^m(\varphi)\langle N_0|H_{TL}|N_0'\rangle=0\label{eq:schrodinger}
\end{align}
where we introduced the Hamiltonian of the transmission line $H_{TL}=\sum_q \hbar \omega_q a_q^\dagger a_q$. 
We further integrate~\eqref{eq:schrodinger} over $u^n_{N_0}$ and find
\begin{align}
        &\left[E_n(N_0)-E\right]f_n(N_0)+\sum_{m,N_0'}[f_m(N_0')\nonumber\\
        &\times\left(\int_0^{2\pi}\frac{d\varphi}{2\pi}u_{N_0}^n(\varphi)^*u_{N_0'}^m(\varphi)\right)\langle N_0|H_{TL}|N_0'\rangle]=0.\label{eq:schrodinger2}
\end{align}
In the limit $E_J\gg E_C$, $u_{N_0}^n(\varphi)$ can be approximated by
\begin{equation}
    u_{N_0}^n(\varphi)=W_n(\varphi)e^{-iN_0\varphi}.
\end{equation}
The other terms in expression~\eqref{eq:Wannier} do not contribute since they are strongly localized close to $2\pi p$ for $p\neq 0$, whereas the integrals are limited to the interval $[0,2\pi]$. At low energy, the set of integral equations~\eqref{eq:schrodinger2} can also be obtained from the effective Hamiltonian
\begin{align}
    H_{\rm dual}=&\,\frac{E_J\hat{\varphi}^2}{2}+4 E_C \hat{N}^2-\Delta\cos 2\pi\hat{N}_0 \nonumber\\[1mm] &+e^{i\hat{N}_0\hat{\varphi}}H_{TL}e^{-i\hat{N}_0\hat{\varphi}}.\label{eq:dual}
\end{align}
The first two terms simply describe an harmonic oscillator with the resonance frequency $\omega_P$, corresponding to an isolated transmon with no external charge. The coupling to the transmission line is entirely carried by the last term. After a gauge transformation, $\tilde{H}_{\rm dual}= U  H_{\rm dual} U^\dagger$ with $U =e^{-i\hat{N}_0\hat{\phi}} $, we obtain
\begin{align}   \tilde{H}_{\rm dual}=\,&E_J\frac{\hat{\varphi}^2}{2}+4E_C(\hat{N}-\hat{N}_0)^2 \nonumber\\[1mm]& -\Delta \cos 2\pi \hat{N}_0+\sum_q \hbar \omega_q a_q^\dagger a_q.\label{eq:hamiltonian_dual} 
\end{align}
The Hamiltonian becomes quadratic for $\Delta = 0$ and describes a transmission line coupled at its end to an LC circuit, as shown in Fig.~\ref{fig:dual}. In section~\ref{sec:BSG} and appendix~\ref{app:gauge-transform}, the Hamiltonian of the transmission line shunted by a capacitance has been diagonalized and the circuit operators re-expressed with new creation and annihilation operators $b_q$ and $b^\dagger_q$, employing input-ouput theory~\cite{clerk2010,beltran2010development}. A similar diagonalization is applied here with the transmission line closed by $C$ in parallel with $L_J$. It leads to the Hamiltonian Eq.~\eqref{eq:hamiltonian_dual2} where the new bosonic fields are related to $\hat N_0$ via Eq.~\eqref{eq:charge_mode}.

\begin{figure}[t!]
	\centering
	\includegraphics[width=\columnwidth]{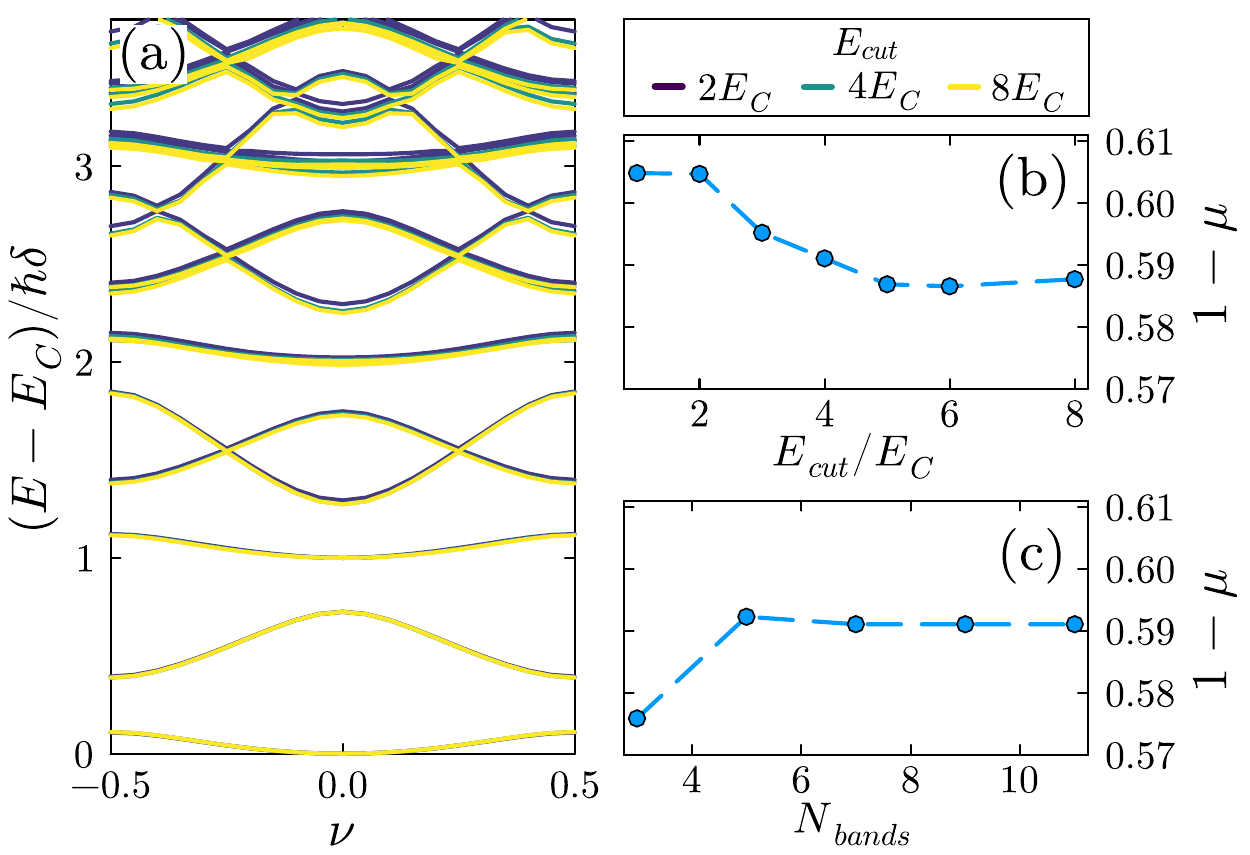}
	\caption{Convergence of the exact diagonalization for $E_J=E_C$, $\hbar\Omega=4E_C$ and $\hbar\delta=0.5E_C$. (a) Spectra for different values of the photonic energy cutoff $E_{cut}$. (b-c) As in Fig.~\ref{fig:ED_fitted_mob}, the mobility is extracted from fitting the lowest bands to Eq.~\eqref{eq:cft-bands}. The convergence is shown with  $E_{cut}$ (b) and $N_{bands}$ (c).}
\label{fig:ED_convergence}
\end{figure}

\section{Exact diagonalization}\label{app:exact-diag}
We performed exact diagonalization of Hamiltonian~\eqref{hamiltonian} in the charge gauge, with the environment represented as a finite-size lumped-element transmission line. In this gauge the Hamiltonian is manifestly invariant under translations of $2\pi$ of the junction's phase. This allows to apply Bloch's theorem on the full Hamiltonian, whose spectrum is hence composed of bands, analogously to the isolated transmon discussed in Appendix~\ref{app:transmon}. Exact diagonalization on this system was already performed in~\cite{Giacomelli23} for various impedances and focusing on the low-lying energy bands. In that work, the charge gauge Hamiltonian was represented on the basis composed by eigenstates of the non-interacting Hamiltonian, i.e. states that are a tensor product of eigenstates of the bare junction Hamiltonian~\eqref{eq:ham_JJ} and of Fock states for the environmental photons.

Here instead, we chose to represent the Hamiltonian in the \textit{polaron frame} basis. This basis is obtained by applying the unitary transformation~\eqref{unitary} to the states $\ket{N}\otimes_q\ket{n_q}$, i.e. the tensor product of junction's charge eigenstates and environmental photons' number states. As discussed in~\cite{Giacomelli23}, this unitary transformation is a charge-dependent displacement operation for each photonic mode. The polaron frame basis is hence built of junction's charge eigenstates and displaced number states for the photons. The Hamiltonian matrix elements between these states can be expressed analytically in terms of associated Laguerre polynomials~\cite{Giacomelli23}.

For each value of the quasi-charge $\nu$, we numerically compute the first $N_{bands}$ levels of the isolated junction Hamiltonian and build the corresponding discrete-spectrum charge operator, of which we find the eigenvectors. We then solve the capacitively terminated transmission line for the modes' frequencies $\omega_q$ and the corresponding couplings to the junction. The numerical basis is composed of all states whose photonic occupation numbers $n_q$ are such that $\sum_q \hbar\omega_q n_q<E_{cut}$. Both $N_{bands}$ and $E_{cut}$ are numerical parameters with respect to which we need to assure convergence.

Analogous numerical cutoffs are present for the non-interacting basis. The representation used here however provides a much faster convergence with respect to $E_{cut}$, i.e. with respect to the number of photons, when $\alpha$ is sufficiently larger than zero. In particular, for $\alpha=1$ this allows to compute higher excited bands, such as the ones shown in Fig.~\ref{fig:ED_bands}. The convergence with respect to the two parameters, $N_{bands}$ and $E_{cut}$, is demonstrated in Fig.~\ref{fig:ED_convergence}. Interestingly, qualitatively accurate results are achieved even for $E_{cut}<\hbar\Omega$, as shown in panel~(a). This is because, in the polaron frame, a significant part of the environment is already captured with  zero (displaced) photons. This is further supported by the weak dependence of the fitted mobility on  $E_{cut}$, as depicted in panel~(b).  The lowest band can always be successfully fitted with Eq.~\eqref{eq:cft-bands}, provided the number of bands is sufficiently large.  The fast convergence in $N_{bands}$ is shown in panel~(c).



%

   \clearpage

\onecolumngrid

\vspace{2cm}

\begin{center}
	\textbf{\large Supplementary Material:\\Quantum critical line in the Schmid transition} \\
	[.4cm] Nicolas Paris,$^{1,2}$ Luca Giacomelli,$^1$ Romain Daviet,$^3$ Cristiano Ciuti,$^1$ Nicolas Dupuis,$^2$ and Christophe Mora$^1$\\[.1cm]
	{\itshape $^1$Universit\'e Paris Cit\'e, CNRS,  Laboratoire  Mat\'eriaux  et  Ph\'enom\`enes  Quantiques, 75013  Paris,  France} \\
	{\itshape $^2$Sorbonne Universit\'e, CNRS, Laboratoire de Physique Th\'eorique de la Mati\`ere Condens\'ee, LPTMC, F-75005 Paris, France}\\
	{\itshape $^3$Institut f\"ur Theoretische Physik, Universit\"at zu K\"oln, D-50937 Cologne, Germany}\\
\end{center}

\setcounter{section}{0} 
\setcounter{equation}{0}
\setcounter{figure}{0}
\setcounter{table}{0}
\renewcommand{\theequation}{S\arabic{equation}}
\renewcommand{\thefigure}{S\arabic{figure}}
\renewcommand{\bibnumfmt}[1]{[S#1]}
\renewcommand{\citenumfont}[1]{S#1}
\renewcommand{\thesection}{\Roman{section}}
\renewcommand{\theHfigure}{S\arabic{figure}}
\renewcommand{\theHequation}{S\arabic{equation}}
\renewcommand{\thesubsection}{\Roman{subsection}}
\renewcommand{\thesubsubsection}{\arabic{subsubsection}}

	\subsection{Hamiltonian of the shunted Josephson junction in different gauges}
	In this section, we bring together the most usual gauge Hamiltonians used in the context of a Josephson junction shunted by a finite transmission line with $N_m$ modes. The limit $N_m\rightarrow \infty$ must be taken to recover the continuum limit studied in the main text.
	
\begin{figure}
		\centering
		\includegraphics[width=0.5\textwidth]{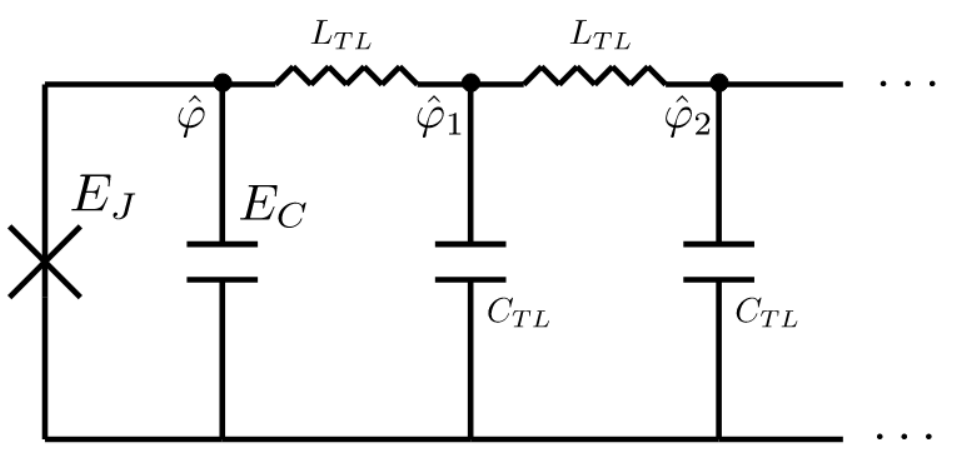}
		\caption{Transmission line terminated by a Josephson junction.}
		\label{fig:TL}
	\end{figure}

	In a classical approach, the Lagrangian of the circuit of Fig.~\ref{fig:TL} can be written in terms of independent node variables~\cite{devoret1995}:
	\begin{equation}
		\mathcal{L}=\frac{1}{8e^2} \left[C\dot{\varphi}^2+C_{TL}\sum_{j=1}^{N_m}\dot{\varphi}_j^2 \right]-\frac{1}{8e^2L_{TL}}\left[(\varphi_1-\varphi)^2+\sum_{j=1}^{N_m}(\varphi_j-\varphi_{j-1})^2\right]+E_J\cos \varphi,
	\end{equation}
	where $E_J$ is the Josephson energy, $E_C= e^2/2 C$ the charging energy of the junction and $\varphi$ the Josephson compact phase. The oscillators of the transmission line have a capacitance $C_{TL}$ and an inductance $L_{TL}$. The flux through each inductor is denoted by $\varphi_j$ with $j\in \{1,\hdots, N_m\}$ and we write $\varphi_0=0$ for convenience.
	The circuit can be quantized by switching to the Hamiltonian formalism:
	\begin{equation}\label{eq:flux}
		H_{\rm flux}=4E_C\hat{N}^2-E_J\cos \hat\varphi+\frac{2e^2}{C_{TL}}\sum_{j=1}^{N_m}\hat n_j^2+\frac{1}{8e^2L_{TL}}\left[(\hat\varphi_1-\hat\varphi)^2+\sum_{j=1}^{N_m}(\hat \varphi_j-\hat\varphi_{j-1})^2\right],
	\end{equation}
	where $[\hat\varphi_j,\hat n_j]=i$ and $[\hat\varphi,\hat N]=i$. The operator $\hat{n}_j$ is the charge conjugated to the flux $\hat \varphi_j$ and $\hat N$ is the number of Cooper pairs transferred across the junction. We define $\hat\varphi_0=0$ for convenience. Equation~\eqref{eq:flux} is the flux gauge Hamiltonian. The coupling between the junction and the bath is quadratic in $\hat\varphi$. One can naively think that it tends to localize the phase $\hat\varphi$, promoting a superconducting state. Nevertheless, this interpretation is simply a shortcut induced by the flux gauge expression of the Hamiltonian: one cannot ignore the effect of the other terms in $H_{\rm flux}$.
	
	Let us make contact with the Hamiltonian in the main text by defining the total flux in the transmission line $\hat{\phi}_0$ and the total charge displaced from the transmission line to the junction $\hat{N}_0$:
	\begin{equation}
		\begin{split}
			&\hat \phi_0=\frac{1}{N_m}\sum_{j=1}^{N_m}\hat\varphi_j\\
			&\hat N_0=\sum_{j=1}^{N_m}\hat n_j
		\end{split}
	\end{equation}
	Note that they are conjugate: $[\hat \phi_0,\hat N_0]=i$.
	The charge gauge Hamiltonian is obtained from~\eqref{eq:flux} with the unitary transformation
	$e^{i\hat \varphi \hat N_0}$. It reads:
	\begin{equation}\label{eq:charge}
		H_{\rm charge}=4E_C(\hat N-\hat N_0)^2-E_J\cos \hat\varphi+\left[\frac{2e^2}{C_{TL}}\sum_{j=1}^{N_m}\hat n_j^2+\frac{1}{8e^2L_{TL}}\sum_{j=1}^{N_m}(\hat \varphi_j-\hat\varphi_{j-1})^2\right].
	\end{equation}
	The right-hand term of $H_{\rm charge}$ is diagonalized with the modes $a_q$ in the main text. Considered on its own, it corresponds to the Hamiltonian of a transmission line terminated by a wire since $\hat\varphi_0=0$ in our notations.

	One can then apply the unitary transformation $e^{-i\hat \phi_0 \hat N}$ to obtain the polaron frame Hamiltonian:
	\begin{equation}
		H_{\rm polaron}=4E_C\hat N_0^2-E_J\cos (\hat\phi_0-\hat \varphi)+\left[\frac{2e^2}{C_{TL}}\sum_{j=1}^{N_m}(\hat n_j-\frac{\hat N}{N_m})^2+\frac{1}{8e^2L_{TL}}\sum_{j=1}^{N_m}(\hat \varphi_j-\hat\varphi_{j-1})^2\right].
	\end{equation}
	In the limit $N_m\rightarrow \infty$ one can safely replace $(\hat n_j-\frac{\hat N}{N_m})^2$ by $\hat n_j^2$: the polaronic transmormation does not change the transmission line Hamiltonian when there is an infinite number of modes:
	\begin{equation}
		H_{\rm polaron}=-E_J\cos (\hat\phi_0-\hat \varphi)+\left[4E_C\hat N_0^2+\frac{2e^2}{C_{TL}}\sum_{j=1}^{N_m}\hat n_j^2+\frac{1}{8e^2L_{TL}}\sum_{j=1}^{N_m}(\hat \varphi_j-\hat\varphi_{j-1})^2\right].
	\end{equation}
	This Hamiltonian is considered in Appendix A of the main text. The right-hand term describes a transmission line terminated by a capacitance $C$ and is diagonalized with the modes $b_q$ in the main text. More generally, one can switch to the continuum $N_m\rightarrow \infty$ with the following substitutions:
	\begin{equation}
		\begin{split}
			&L_{TL}\rightarrow l\,dx\\
			&C_{TL}\rightarrow c\,dx\\
			&\hat n_j=\hat n(x_j)\rightarrow  \hat n(x)\,dx\\
			&\hat \varphi_j=\hat \varphi(x_j)\rightarrow \hat \varphi(x)
		\end{split}
	\end{equation}
	with $x_j=j\,dx$. The operators $\hat{\varphi}$ and $\hat{N}$ are unchanged while:
	\begin{equation}
		\begin{split}
			&\hat N_0=\int_0^{+\infty} dx\ \hat{n}(x)\\
			&\hat\phi_0=\int_0^{+\infty}dx \,\hat \varphi(x).
		\end{split}
	\end{equation}
	In this limit, the Hamiltian becomes:
	\begin{equation}
		H_{cont}=4E_C\hat N_0^2-E_J\cos (\hat\phi_0-\hat \varphi)+\int_0^{+\infty}dx\,\left[\frac{2e^2}{c}\hat n(x)^2+\frac{1}{8e^2l}\left(\frac{\partial \hat\varphi(x)}{\partial x}\right)^2\right],
	\end{equation}
	with the usual continuum expression for the transmission line~\cite{Morel21}.
	
	\subsection{\label{app:third_order}Third order $\Gamma_{3}^{(2)}$ in perturbation theory}
\begin{figure}
		\centering
		\includegraphics[width=0.6\textwidth]{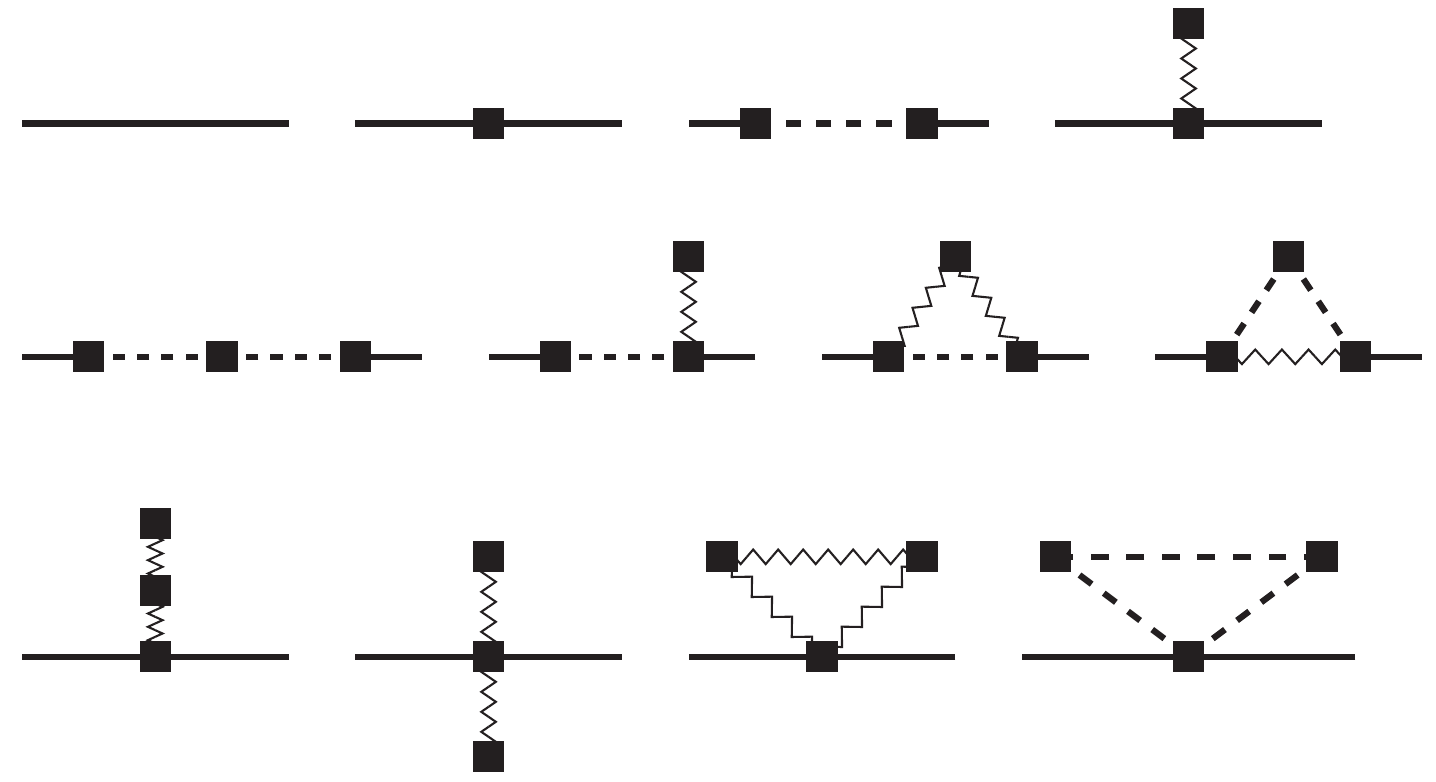}
		\caption{All the diagrams contributing to the renormalization of $\Gamma^{(2)}$ up to third order.}
		\label{fig:order_3}
	\end{figure}

	At order 3 in $\tilde{E}_J^0$, the bare 1-PI two-point vertex reads:
	\begin{equation}
		\Gamma^{(2)}(\omega,\beta_0,\kappa,\tilde{E}_J^0,a)=\sqrt{\omega^2+\kappa^2}+J+\Gamma_{2}^{(2)}+\Gamma_{3}^{(2)}.
	\end{equation}
	where $\Gamma_{2}^{(2)}$ is the second order term given in the main text.
	In this appendix, we compute the logarithmic divergences of the third order term
	\begin{equation}\label{eq:gamma_bare}
		\begin{split}
			\Gamma^{(2)}_{3}&=J^3\beta_0^{-4}\int_{-\infty}^{+\infty} d\tau \int_{-\infty}^{+\infty}d\tau'\,\{ e^{i\omega \tau}\left[\sinh I(\tau)-I(\tau)\right]e^{i\omega \tau'}\left[\sinh I(\tau')-I(\tau')\right]\\
			& -2e^{i\omega \tau}\left[\sinh I(\tau)-I(\tau)\right]\left[\cosh I(\tau')-1\right]\\
			& -e^{i\omega \tau}\sinh I(\tau)\left[\cosh I(\tau')-1\right]\left[\cosh I(\tau-\tau')-1\right]\\
			& +e^{i\omega \tau}\left[\cosh I(\tau)-1\right]\sinh I(\tau')\sinh I(\tau-\tau')\\
			& +\frac{3}{2}\left[\cosh I(\tau)-1\right]\left[\cosh I(\tau')-1\right]\\
			& +\frac{1}{2}\left[\cosh I(\tau)-1\right]\left[\cosh I(\tau')-1\right]\left[\cosh I(\tau-\tau')-1\right]\\
			&-\frac{1}{2}\sinh I(\tau)\sinh I(\tau')\sinh I(\tau-\tau')\}.
		\end{split}
	\end{equation}
	All the contributing Feynman diagrams are drawn in Fig.~\ref{fig:order_3}. We recovered exactly the same diagrams as in~\cite{amit1980renormalisation}. It is not surprising since the interaction term is a cosine in both cases.

	The only sources of divergences are $J$ and the integrals. The former is already taken into account in the renormalisation constant $Z_{tadpole}$ and we need not bother with it any longer. We deal with the integrals as for $\Gamma_{2}^{(2)}$ by expanding the exponential in powers of $\omega$. Note that once again, there is no term proportional to $\omega$ for the sake of parity. Therefore, we are left with $\omega-$independent terms and $\omega^2$-dependent terms. The higher order terms do not diverge in the UV.
	
	\subsubsection{$\omega^2$-dependent contribution}
	One can extract the terms proportional to $\omega^2$ in~\eqref{eq:gamma_bare} :
	\begin{equation}
		\begin{split}
			\Gamma^{(2)}_{3,\omega^2}&=J^3\beta_0^{-4}\omega^2\int_{-\infty}^{+\infty} d\tau \int_{-\infty}^{+\infty}d\tau'\,(-\tau^2)\left[\sinh I(\tau)-I(\tau)\right]\left[\sinh I(\tau')-I(\tau')\right]\\
			&+\tau^2\left[\sinh I(\tau)-I(\tau)\right]\left[\cosh I(\tau')-1\right]\\
			&+\frac{\tau^2}{2}\sinh I(\tau)\left[\cosh I(\tau')-1\right]\left[\cosh I(\tau-\tau')-1\right]\\
			&-\frac{\tau^2}{2}\left[\cosh I(\tau)-1\right]\sinh I(\tau')\sinh I(\tau-\tau').
		\end{split}
	\end{equation}
	Retaining only the potentially divergent terms, one finds:
	\begin{equation}\label{eq:omega2_term}
		\Gamma^{(2)}_{3,\omega^2}=J^3\beta_0^{-4}\omega^2\int_{-\infty}^{+\infty} d\tau \int_{-\infty}^{+\infty}d\tau'\,\frac{\tau^2}{2}\left[\sinh I(\tau)\cosh I(\tau')\cosh I(\tau-\tau')-\cosh I(\tau)\sinh I(\tau')\sinh I(\tau-\tau')\right].
	\end{equation}
	where the '$=$' sign is to be taken as an identification regarding the UV divergences. We will use the same kind of identification all along this appendix. Up to now, our calculations are very similar to those presented in~\cite{amit1980renormalisation}. However, at this point, we can simplify~\eqref{eq:omega2_term} further for symmetry reasons. Extracting the UV dangerous contributions in~\eqref{eq:omega2_term} yields:
	\begin{equation}\label{eq:omega2_dv}
		\Gamma^{(2)}_{3,\omega^2}=J^3\beta_0^{-4}\omega^2\int_{-\infty}^{+\infty} d\tau\int_{-\infty}^{+\infty} d\tau'\,\frac{\tau^2}{8}\left[2 e^{I(\tau)+I(\tau')-I(\tau-\tau')}-e^{-I(\tau)+I(\tau-\tau')+I(\tau')}\right].
	\end{equation}
	The second contribution can be rewritten as:
	\begin{equation}
		\begin{split}
			-\int_{-\infty}^{+\infty} d\tau \int_{-\infty}^{+\infty}d\tau'\,\frac{\tau^2}{8}e^{-I(\tau)+I(\tau-\tau')+I(\tau')}&=-\int d\tau d\tau'\,\frac{(\tau+\tau')^2}{8}e^{-I(\tau+\tau')+I(\tau)+I(\tau')}\\
			&=-\int d\tau d\tau'\,\frac{(\tau-\tau')^2}{8}e^{-I(\tau-\tau')+I(\tau)+I(\tau')}\\
			&=-\int d\tau d\tau'\,\frac{\tau^2+\tau'^2}{8}e^{-I(\tau-\tau')+I(\tau)+I(\tau')}\\
			&=-\int d\tau d\tau'\,\frac{2 \tau^2}{8}e^{-I(\tau-\tau')+I(\tau)+I(\tau')}.
		\end{split}
	\end{equation}
	Once re-inserted in~\eqref{eq:omega2_dv}, it compensates the first term. Therefore, there is no UV divergence in $\Gamma_{3,\omega^2}^{(2)}$.
	
	\subsubsection{$\omega$-independent contribution}
	The $\omega$-independent contribution in~\eqref{eq:gamma_bare} is a little more complicated. Nevertheless, after a bit of algebra, expanding the hyperbolic sines and cosines, one finds that the potential UV divergence can emerge from:
	\begin{equation}\label{eq:indep}
		\Gamma^{(2)}_{3,indep}=J^3\beta_0^{-4} \frac{1}{8}\int_{-\infty}^{+\infty} d\tau \int_{-\infty}^{+\infty}d\tau'\, \{e^{I(\tau)+I(\tau')-I(\tau-\tau')}-e^{I(\tau)}-e^{I(\tau')}\}.
	\end{equation}
	only. Note that we recover exactly the equation (A.3.1) of~\cite{amit1980renormalisation}.
	We must pay attention to three potentially dangerous regions of integration:
	\begin{eqnarray}\label{eq:regions}
		&&(1)=\{|\tau|<\Delta, |\tau'|>\Delta\}\\
		&&(2)=\{|\tau'|<\Delta, |\tau|>\Delta\}\\
		&&(3)=\{|\tau|<\Delta, |\tau'|<\Delta\},
	\end{eqnarray}
	where $\Delta$ is chosen such that $\Delta \kappa\ll 1$.
	The integral~\eqref{eq:indep} is symmetric under $\tau \leftrightarrow \tau'$, therefore regions $(1)$ and $(2)$ give exactly the same contribution. Let us focus on region $(1)$. When $|\tau|$ is small, $I(\tau)$ can be replaced by its small-$\tau$ expansion. As for $I(\tau-\tau')$, it can be expanded around $I(\tau')$. Last, $e^{I(\tau')}$ does not present any dangerous divergence as long as $|\tau'|>\Delta$. Therefore, in region $(1)$, the potential UV divergence of $\Gamma^{(2)}_{3,indep}$ are fully contained in:
	\begin{equation}
		(\Gamma^{(2)}_{3,indep})^{(1)}=J^3\beta_0^{-4}
		\frac{1}{8}\int_{-\Delta}^{+\Delta} d\tau \int_{-\infty}^{+\infty}d\tau'\, \frac{\tau I'(\tau')+\frac{\tau^2}{2}I''(\tau')}{(\kappa^2c(\tau^2+a^2))^{1/\alpha}}.
	\end{equation}
	The first term vanishes by parity, and the second is convergent for $\alpha>\frac{2}{3}$. Therefore, regions $(1)$ and $(2)$ do not bring any logarithmic divergence.
	We are left with region $(3)$, which is more complicated to deal with. In this region, all the arguments $\tau$, $\tau'$ and $\tau-\tau'$ are small, so that all the propagators can be replaced by their small-$\tau$ behavior:
	\begin{equation}\label{eq:region_3}
		(\Gamma^{(2)}_{3,indep})^{(3)}=J^3\beta_0^{-4}\frac{1}{8(\kappa^2 c)^{1/\alpha}}\int_{-\Delta}^{\Delta} d\tau \int_{-\Delta}^{\Delta}d\tau' \, \left[\left(\frac{((\tau-\tau')^2+a^2)}{(\tau^2+a^2)(\tau'^2+a^2)}\right)^{1/\alpha}-\frac{1}{(\tau^2+a^2)^{1/\alpha}}-\frac{1}{(\tau'^2+a^2)^{1/\alpha}}\right].
	\end{equation}
	When $\alpha=1$, an exact cancellation occurs and $(\Gamma^{(2)}_{3,indep})^{(3)}$ does not diverge when $a$ goes to zero. It definitely proves that the $\beta_J$-function vanishes at third order on the line $\alpha=1$. When $\alpha\neq 1$, $(\Gamma^{(2)}_{3,indep})^{(3)}$ has some logarithmic contribution in $\ln(a)$. It can be dealt with numerically for instance, performing the integral~\eqref{eq:region_3} for several values of $a$ and extracting the $\ln a$ contribution. We chose to expand the integral in powers of $\alpha-1$ and to focus on the first non-vanishing term, which is proportional to $\alpha-1$ and comes with the integral:
	\begin{equation}
		\begin{split}
			(\Gamma^{(2)}_{3,indep})^{(3)}&=\frac{J^3}{8\beta_0^{4}}\left(\frac{1}{\alpha}-1\right)\int_{-\Delta}^{\Delta} d\tau \int_{-\Delta}^{\Delta} d\tau' \, \frac{\ln(\kappa^2c(\tau^2+a^2))}{\kappa^2c(\tau^2+a^2)}+\frac{\ln(\kappa^2c(\tau'^2+a^2))}{\kappa^2c(\tau'^2+a^2)}\\
			&-\frac{((\tau-\tau')^2+a^2)\left[\ln(\kappa^2c(\tau^2+a^2))+\ln(\kappa^2c(\tau'^2+a^2))-\ln(\kappa^2c((\tau-\tau')^2+a^2))\right]}{\kappa^2c(\tau^2+a^2)(\tau'^2+a^2)}.
		\end{split}
	\end{equation}
	Although we did not compute each integral analytically, we found with extremely good numerical accuracy that the third order contribution is given by:
	\begin{equation}\label{eq:result}
		(\Gamma^{(2)}_{3,indep})^{(3)}=-\frac{J^3}{16\kappa^2c}\left(\frac{1}{\alpha}-1\right)\ln(\kappa c^{1/2} a).
	\end{equation}
	In particular, it was checked that the pre-factor of the logarithm does not depend on $\Delta$ when it is small enough compared to $1/\kappa$. Nevertheless, this pre-factor depends on the cut-off procedure. For instance, it varies by a factor 2 between our cut-off prescription and a hard UV cut-off.
	The  divergence in~\eqref{eq:result} can be absorbed as well as the tadpole contribution by defining:
	\begin{equation}
		Z_J=Z_{tadpole}\times Z_{3}
	\end{equation}
	where $Z_3=1+(\frac{1}{\alpha}-1)\frac{\tilde{E}_J^2}{16}\ln(\kappa a)$. The resulting $\beta$-function is given in the main text.

\end{document}